\documentclass[aps,prl,reprint,showpacs,amsmath,amssymb,superscriptaddress]{revtex4-1}
\usepackage{graphicx}

\begin{document}

\title{Deconstructing $1/f$ noise and its universal crossover to non-$1/f$ behavior}

\author{Sebastian A. Diaz}
\email[]{sdiaz@physics.ucsd.edu}
\affiliation{Department of Physics, University of California, San Diego, La Jolla, California 92093, USA}

\author{Massimiliano Di Ventra}
\email[]{diventra@physics.ucsd.edu}
\affiliation{Department of Physics, University of California, San Diego, La Jolla, California 92093, USA}

\date{\today}

\begin{abstract}
Noise of stochastic processes whose power spectrum scales at low frequencies, $f$, as $1/f$ appears in such diverse systems that it is considered universal. However, there have been a small number of instances from completely unrelated fields, e.g., the fluctuations of the human heartbeat or vortices in superconductors, in which power spectra have been observed to cross over from a $1/f$ to a non-$1/f$ behavior at even lower frequencies. Here, we show that such crossover must be universal, and can be accounted for by the memory of initial conditions and the relaxation processes present in any physical system. When the smallest frequency allowed by the experimental observation time, $\omega_{obs}$, is larger than the smallest relaxation frequency, $\Omega_{min}$, a $1/f$ power spectral density is obtained. Conversely, when $\omega_{obs}<\Omega_{min}$ we predict that the power spectrum of \emph{any} stochastic process should exhibit a crossover from $1/f$ to a different, integrable functional form provided there is enough time for experimental observations. This crossover also provides a convenient tool to measure the lowest relaxation
frequency of a physical system.
\end{abstract}


\maketitle

%%%%%%%%%%%%%%%%%%%%%%%%%%%%%%%%%%%%%%%%
%%%%%%%%%%%%%%%%%%%%%%%%%%%%%%%%%%%%%%%%
In 1925 J. B. Johnson published his results on investigations of voltage drop fluctuations across vacuum tubes \cite{Johnson:1925}. He found discrepancies with the theory of Schottky \cite{Schottky:1918,Schottky:1922} for the \emph{small-shot effect}---nowadays known as \emph{shot noise}---, especially at low frequencies. Shortly after, Schottky provided an explanation for Johnson's findings based on a microscopic model specific to the physics of the cathode surface \cite{Schottky:1926}. Furthermore, comparing the fluctuations in voltage to the variations in brightness of a source of light---known as flicker---he named this phenomenon the \emph{flicker effect}. In modern parlance this is referred to as \emph{flicker noise} or $1/f$ \emph{noise}.

Strikingly, low-frequency behavior similar to the flicker effect in vacuum tubes has been observed in a plethora of phenomena from such different and unrelated fields as solid-state physics \cite{DuttaHorn:1981,VossClarke:1976}, astronomy \cite{Press:1978,Lyubarskii:1997}, physiology \cite{Derksen:1966,Neumcke:1978,Peng:1995}, meteorology \cite{Yano:2001,Yano:2004,Fraedrich:2009}, geophysics \cite{Wunsch:1972,Wunsch2:1972,Mao:1999}, geology \cite{Holliger:1996}, economics \cite{Bonanno:2000}, and music \cite{VossClarke:1975,VossClarke:1978}, to name just a few. Common to all of them is the fact that the associated \emph{power spectral density} scales as the inverse of a power of the frequency at low frequencies, namely $1/f^{\alpha}$, with $\alpha$ typically ranging from 1 to 2 \cite{DuttaHorn:1981}.

The ubiquitousness of $1/f$ noise has stimulated researchers from different fields to study and try to explain this intriguing phenomenon, but no general consensus on its physical origin has been reached yet. Nevertheless, all the attempts found in the literature seem to fall within two categories: those that aim at explaining a particular set of experiments by means of specific models---as Schottky first did \cite{Schottky:1926}---, and those that aim at generating a theory that may encompass the whole body of observed phenomena exhibiting $1/f$ noise. For instance, one well-known attempt of the latter kind bears the name of ``self-organized criticality'' \cite{Bak:1987,Bak:1988}. According to this theory, some spatially extended and dissipative dynamical systems naturally evolve toward a critical state where there are no characteristic time or length scales. The hallmarks of such a state being the presence of $1/f$ noise and the emergence of fractal (scale-invariant) structures in space. This appealing idea was met with enthusiasm, but also raised controversy in the literature once put to experimental and numerical tests \cite{Anderson:1996,Frigg:2003}.

Even more puzzling is the fact---not much appreciated or discussed in the literature---that in some (although not many) instances, the $1/f$ law gives way to a non-$1/f$ dependence if the frequency is pushed to even lower experimental observation values. This crossover appears again in cases that could not be farther from each other, e.g., voltage fluctuations in superconducting thin films \cite{ClarkeHsiang:1976}, fluctuations in the membrane potential of nerve cells \cite{Verveen:1965,Verveen:1968}, voltage fluctuations due to transport of magnetic flux vortices in type-II superconductors \cite{Gurp:1968,Clem:1981}, critical-current fluctuations in Josephson junctions \cite{Anton:2012}, and fluctuations in the inter-beat interval of the human heart \cite{Peng:1995}. However, it does not seem to be present in the majority of experiments.

Why then some, and not all, stochastic processes show a crossover from $1/f$ to non-$1/f$ behavior? Is this crossover specific to just these cases, or we should be able to observe it in \emph{all} cases, provided we could push the experimental observation frequencies to extremely low values? And if so, how low must these frequencies be?

In this Letter, we will explicitly show that $1/f$ noise arises from two factors that are inextricable from any fluctuating observable measured in experiments: its \emph{initial conditions} and \emph{relaxation processes}. By employing only these two physical facts we derive in a transparent and completely general case  the $1/f$ dependence of the power spectrum. Most importantly, we show that this dependence is limited to frequencies larger than the smallest relaxation frequency, $\Omega_{min}$, of the system---\emph{irrespective}  of the physical mechanisms that lead to such relaxation. When the experimental observation frequency is smaller than $\Omega_{min}$ a completely different functional form of the noise power emerges, which converges to a finite value (possibly zero) in the limit of zero frequency. We thus predict that this crossover to a non-$1/f$ behavior should be observable in \emph{all} stochastic processes provided the experimental time scale for observation could be pushed to the slowest relaxation times. Unfortunately, for many physical systems, the smallest relaxation frequencies may correspond to relaxation times that are extremely long to be probed in any practical experiment. However, in some physical systems, such a crossover can be observed quite easily. As an example, we show this crossover in the most familiar case of a Brownian particle in 1-D, but there are other systems, such as glassy systems, where the relaxation times span a very broad spectrum \cite{Refregier:1987}. We suggest these systems as the ones where a systematic analysis of our predictions could be tested, in addition to those we already mentioned.

%%%%%%%%%%%%%%%%%%%%%%%%%%%%%%%%%%%%%%%%
%%%%%%%%%%%%%%%%%%%%%%%%%%%%%%%%%%%%%%%%
\emph{$1/f$ noise facts}.---
Let us begin by clarifying explicitly what is meant by $1/f$ noise. A noisy, time-fluctuating physical variable $A(t)$---a stochastic process---is said to exhibit $1/f$ noise when the power spectral density of its fluctuations $\delta A(t)$, defined by
\begin{equation}\label{eq:PSD}
S(\omega)\equiv\langle|\delta\hat{A}(\omega)|^2\rangle,
\end{equation}
behaves as $1/\omega^\alpha$ for small frequencies, typically with $1\leq\alpha\leq2$. Above, $\delta\hat{A}(\omega)$ denotes the Fourier transform of $\delta A(t)$, and the symbol $\langle \cdots \rangle$ denotes the ensemble average over the stochastic realizations of the process. The information contained in $S(\omega)$ is the answer to the question: How much of the fluctuation of $A(t)$ is, on average, at a given frequency $\omega$? For its ubiquitousness and the unusual properties to be discussed below, in this paper we shall be concerned with the $\alpha=1$ case, but our conclusions are valid for other exponentials as well.

One of the properties of $1/f$ noise that first stands out is the frequency range where it takes place. Astonishingly, experiments show that $S(\omega)\approx 1/\omega$ persists over many frequency decades. This observation plus the realization that it all happens at low frequencies, corresponding to long observational times, hint at some sort of ``encoding'' of a long-time memory effect. In other words, past states appear to have a strong influence on the present state. Thus, initial conditions seem to play an important role in the low-frequency noise properties of these systems.

The integral of $S(\omega)$ represents the total power stored in the fluctuations of the stochastic process over the chosen interval of integration. Therefore, if the functional form $1/\omega$ extends to the origin, the total power over any interval from the origin would be infinite---a physical impossibility. Two potential solutions have been suggested in the literature to overcome this ``infrared catastrophe''. The first attempted solution postulates the existence of a minimum frequency below which the power spectral density transforms into an integrable function \cite{VanDerZiel:1950}. There is not much evidence for the existence of this elusive minimum frequency, perhaps because of the experimental challenge to resolve such low frequencies. Nevertheless, as we have already mentioned, a handful of experiments have reported a change in the $1/\omega$ functional form \cite{ClarkeHsiang:1976,Verveen:1965,Verveen:1968,Gurp:1968,Clem:1981,Peng:1995,Anton:2012} that we argue are proof of its existence. The second potential solution put forward the idea of acknowledging the non-stationary character of the process yielding flicker noise \cite{Mandelbrot:1967}. This indeed seems appropriate by considering the long-term memory property mentioned before: it appears questionable to treat as stationary a process whose present state is strongly influenced by its past dynamics.
\begin{figure}[t]
\centering
\includegraphics[width=\columnwidth]{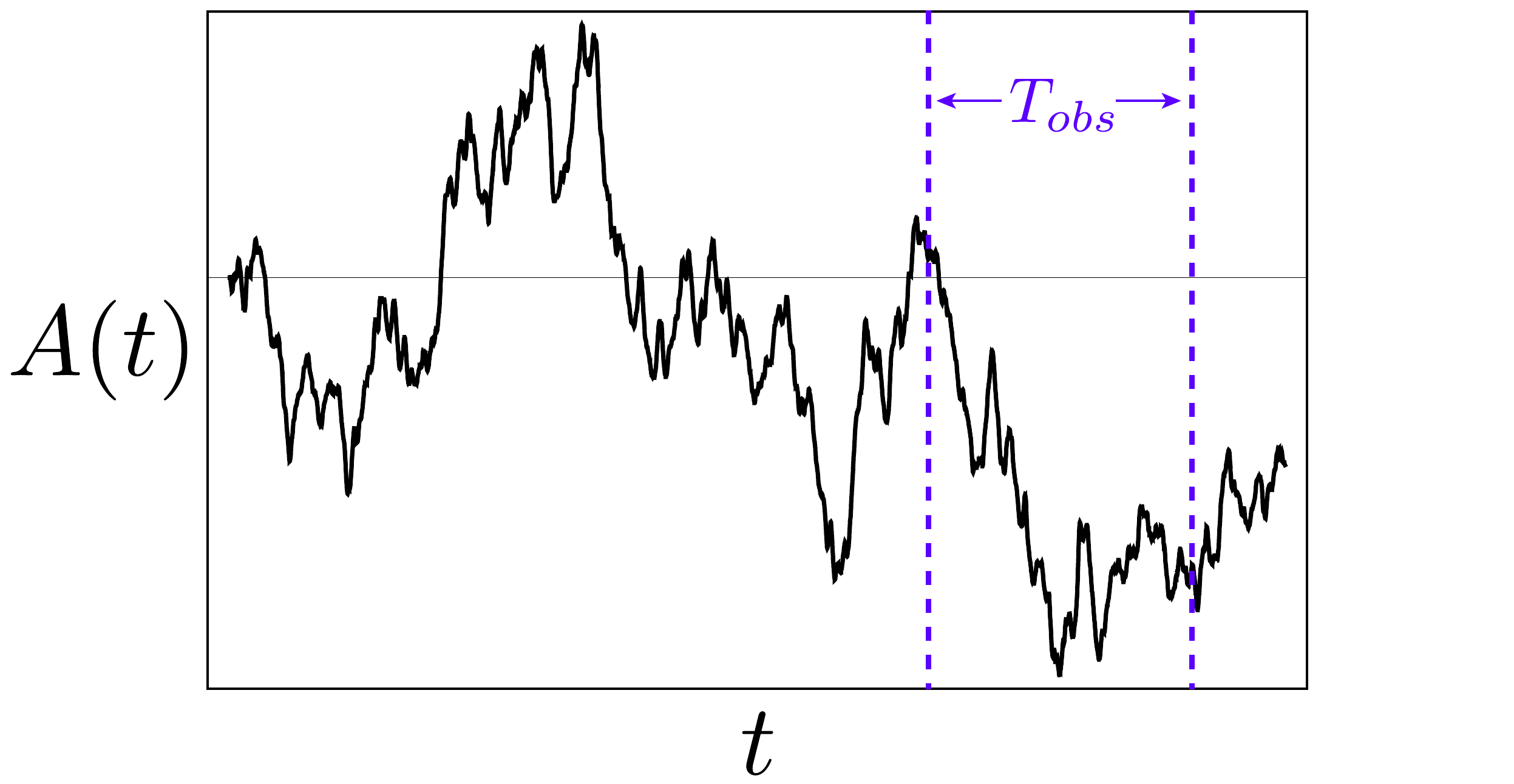}
\caption{(color online). A typical stochastic process $A(t)$. Experimental data for a stochastic process $A(t)$ that started at $t=0$ is typically taken within a window of time of length $T_{obs}$ much shorter than the time elapsed since the beginning of the process.}
\label{fig:TimeWindow}
\end{figure}

%%%%%%%%%%%%%%%%%%%%%%%%%%%%%%%%%%%%%%%%
%%%%%%%%%%%%%%%%%%%%%%%%%%%%%%%%%%%%%%%%
\emph{Initial conditions and relaxation}.---
Therefore, our point of view is that the above feature, namely, memory of \emph{initial conditions} and hence non-stationarity should be taken into account from the outset in any theory aiming at explaining flicker noise. In fact, in Nature there is no such a thing as an \emph{ideal} stationary stochastic
process \cite{DiVentra:2008}: although we may imagine running an experiment forever into the future---as unpractical as this may sound---it is obvious that every process must have started from some initial condition. Notwithstanding, a major assumption that greatly facilitates the analysis of noisy data is to regard the stochastic process at hand as stationary. This working approximation, at first glance, seems harmless mainly because many well established results such as Johnson-Nyquist noise and shot noise \cite{DiVentra:2008} have been obtained relying on its use. However, as anticipated above, it looks like this approximation is not appropriate for $1/f$ noise investigations \cite{Keshner:1982}, precisely because low frequencies---related to long-time correlations---are involved, whereas for other types of noise the focus has been on higher frequencies---related to short-time correlations. Thus, initial conditions and long-term memory effects are not that critical for other types of noise where the stationarity approximation has been employed successfully.

There is also an extra physical requirement that needs to be taken into account. Whenever energy/momentum can be exchanged through interactions with a large number of degrees of freedom, \emph{relaxation} processes take place. The systems exhibiting $1/f$ noise are no exception.

Let us then model the spontaneous fluctuations of a stochastic process $A(t)$ starting at some arbitrary time $t=0$ (an example of this is represented in Fig. \ref{fig:TimeWindow}) by the following superposition of relaxation processes
\begin{equation}
\delta A(t)=\sum_i A_i^0 e^{-\Omega_i(t-t_i)}\theta(t-t_i).
\end{equation}
Note that such a representation has also been employed by Agu \cite{Agu:1976} to derive a $1/f$ power spectral density, but his analysis has not been pushed to our conclusions.

\begin{figure}[t]
\centering
\includegraphics[width=\columnwidth]{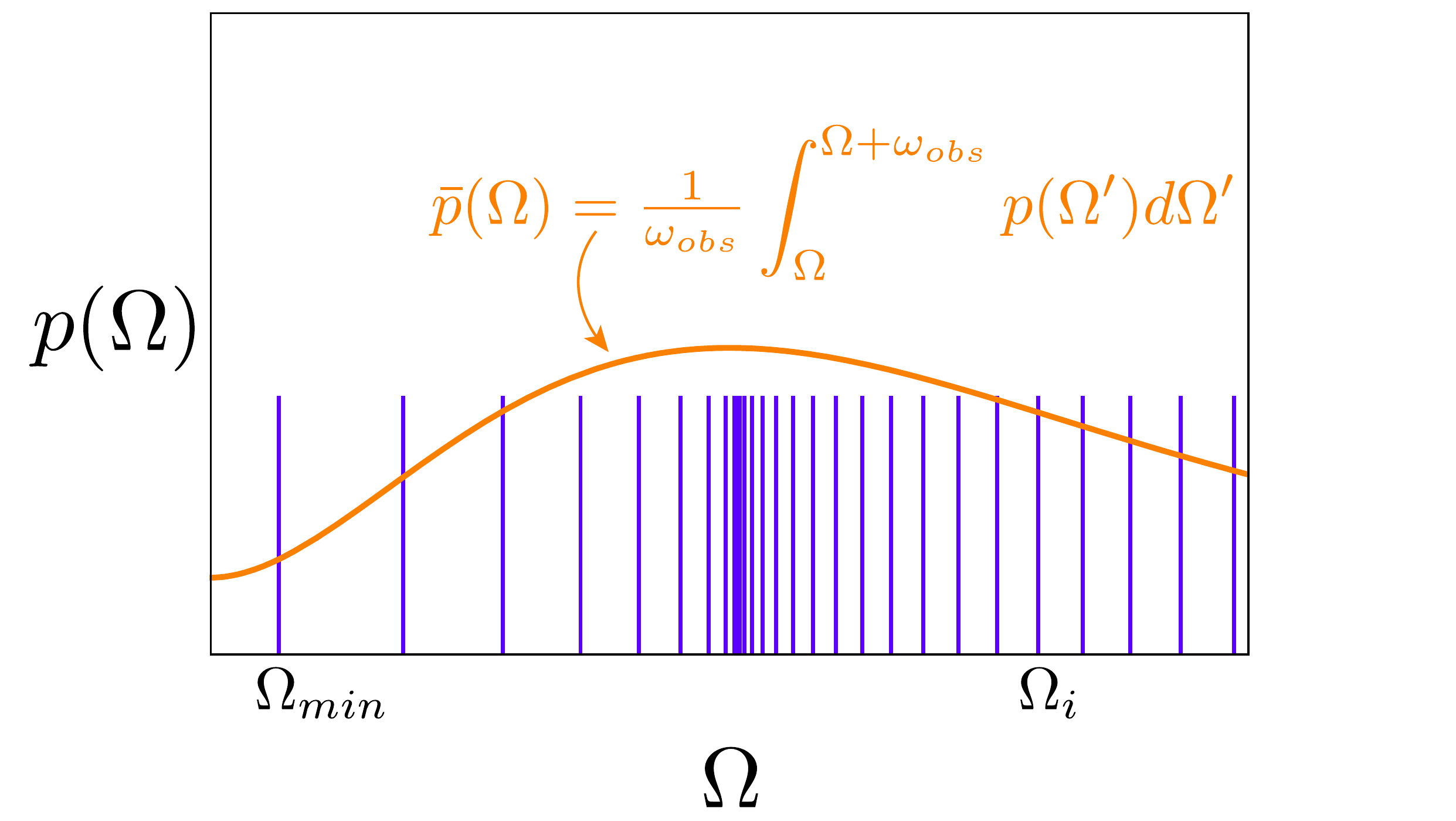}
\caption{(color online). Density of relaxation frequencies $p(\Omega)$. The relaxation frequencies $\Omega_i$ (blue) form a discrete set, hence the exact $p(\Omega)$ is discontinuous. Because the frequency resolution is determined by $\omega_{obs}=2\pi/T_{obs}$, experiments measure an average density $\bar{p}(\Omega)=\tfrac{1}{\omega_{obs}}\int_{\Omega}^{\Omega+\omega_{obs}}p(\Omega')d\Omega'$. A continuous average is obtained when $\omega_{obs}\gg\delta\Omega$, with $\delta\Omega$ being the average distance between successive $\Omega_i$'s. In the schematic plot the orange curve represents such average.}
\label{fig:RelaxFreqDens}
\end{figure}

The $i$-th relaxation process kicks in at time $t_i$ with an amplitude $A_i^0$ and decays with a relaxation frequency $\Omega_i$. Before the time $t_i$ the $i$-th relaxation process was zero as enforced by the Heaviside step function $\theta(\cdot)$. The initial amplitudes $A_i^0$ and the initial times $t_i$ are assumed to be independent random variables. On physical grounds, we know that the average amplitude of fluctuations at the time they kick in must vanish, namely $\langle A_i^0\rangle=0$. On the other hand  $\langle|A_i^0|^2\rangle$ assumes some finite value which could be different for each relaxation process. However, although not necessary for our conclusions, the calculations
are analytically tractable if we assume $\langle|A_i^0|^2\rangle=\Delta^2$,  with $\Delta>0$. Naturally, $\forall i$, $t_i\geq 0$ and $\Omega_i >0$. The $\Omega_i$'s are not all the same. In fact, there exists a distribution of relaxation frequencies modeled by some density function $p(\Omega)$ (see Fig. \ref{fig:RelaxFreqDens}).

%%%%%%%%%%%%%%%%%%%%%%%%%%%%%%%%%%%%%%%%
%%%%%%%%%%%%%%%%%%%%%%%%%%%%%%%%%%%%%%%%
\emph{$1/f$ noise}.---
We now proceed to compute the power spectral density by its definition \eqref{eq:PSD}. The Fourier transform of the fluctuation is
\begin{equation}
\delta\hat{A}(\omega)=\int_{-\infty}^{\infty} \delta A(t)e^{-i\omega t} dt=\sum_i\frac{A_i^0 e^{-i\omega t_i}}{i\omega+\Omega_i}.
\end{equation}
Then, the square of the absolute value of $\delta\hat{A}(\omega)$ is computed as follows
\begin{eqnarray}
\!\!\!\!|\delta\hat{A}(\omega)|^2&=&\delta\hat{A}(\omega)\times\delta\hat{A}(\omega)^* \nonumber\\
&=&\sum_i\frac{|A_i^0|^2}{\omega^2+\Omega_i^2} + \sum_{i\neq j}\frac{A_i^0 A_j^{0*} e^{-i\omega(t_i-t_j)} }{(i\omega+\Omega_i)(-i\omega+\Omega_j)}.
\end{eqnarray}
Ensemble averaging $|\delta\hat{A}(\omega)|^2$ and recalling that the initial amplitudes and times were assumed to be independent random variables we get
\begin{equation}
\langle|\delta\hat{A}(\omega)|^2\rangle=\sum_i\frac{\langle|A_i^0|^2\rangle}{\omega^2+\Omega_i^2} + \sum_{i\neq j}\frac{\langle A_i^0\rangle \langle A_j^{0*} \rangle\langle e^{-i\omega(t_i-t_j)}\rangle }{(i\omega+\Omega_i)(-i\omega+\Omega_j)}.
\end{equation}
By virtue of the demand $\langle A_i^0\rangle=0$, the double sum containing the interference terms vanishes. We are then left with only
\begin{equation}
\langle|\delta\hat{A}(\omega)|^2\rangle=\sum_i\frac{\Delta^2}{\omega^2+\Omega_i^2},
\end{equation}
where $\langle|A_i^0|^2\rangle=\Delta^2$ was used. Invoking the very definition of the Dirac delta function we can write
\begin{equation}
\langle|\delta\hat{A}(\omega)|^2\rangle=\Delta^2\int_0^\infty d\Omega\frac{p(\Omega)}{\omega^2+\Omega^2},
\end{equation}
with $p(\Omega)\equiv\sum_i\delta(\Omega-\Omega_i)$ being the anticipated density of relaxation frequencies. At this stage of the calculation it is convenient to introduce the following function
\begin{equation}
\mathcal{K}_\omega(\Omega)\equiv\frac{2\omega}{\pi(\omega^2+\Omega^2)},
\end{equation}
which has the property $\mathcal{K}_\omega(\Omega)\xrightarrow{\omega\to 0}\delta(\Omega)$. Provided that the density of relaxation frequencies be analytic about zero, and that $p(0)$ be finite and non-vanishing, it then follows
\begin{widetext}
\begin{equation}\label{eq:1/f}
S(\omega)=\Delta^2\int_0^\infty d\Omega\frac{p(\Omega)}{\omega^2+\Omega^2}=\frac{\pi}{2\omega}\Delta^2\int_0^\infty d\Omega\;p(\Omega)\mathcal{K}_\omega(\Omega)\xrightarrow{\omega\to 0}\frac{\pi}{2\omega}\Delta^2p(0).
\end{equation}
\end{widetext}

%%%%%%%%%%%%%%%%%%%%%%%%%%%%%%%%%%%%%%%%
%%%%%%%%%%%%%%%%%%%%%%%%%%%%%%%%%%%%%%%%
\emph{Crossover to non-$1/f$ behavior}.---
We have thus derived a power spectral density proportional to the inverse of frequency in the limit of small frequencies. A statistical measure of the initial conditions, $\Delta^2$, appears in this result. However, equation \eqref{eq:1/f} contains $p(0)$, the density of zero relaxation frequencies, or equivalently, of \emph{infinite} relaxation time. Although it is very appealing that the $S(\omega)$ we derived for low frequencies has $p(0)$ as a coefficient, it is unphysical to have relaxation processes that never relax. As explained in Fig. \ref{fig:RelaxFreqDens}, $p(\Omega)$ should be instead understood as
\begin{equation}
\bar{p}(\Omega)=\tfrac{1}{\omega_{obs}}\int_{\Omega}^{\Omega+\omega_{obs}}p(\Omega')d\Omega', \label{pav}
\end{equation}
the average that accounts for the smallest allowed observational frequency $\omega_{obs}$. A similar rationale is behind $\mathcal{K}_\omega(\Omega)$ and its $\omega\to 0$ limit. As $\omega$ tends to $\omega_{obs}$, $\mathcal{K}_\omega(\Omega)$ tends to a function whose support is $[0, \omega_{obs})$. When such support contains $\Omega_{min}$, that is to say, when $\Omega_{min}<\omega_{obs}$, $p(\Omega)$ cannot be resolved in the interval $[0, \Omega_{min}]$ and $p(0)>0$ should be regarded as $\bar{p}(\Omega_{min})$. Thus, $1/f$ noise is obtained in this case.

\begin{figure}[t]
\centering
\includegraphics[width=\columnwidth]{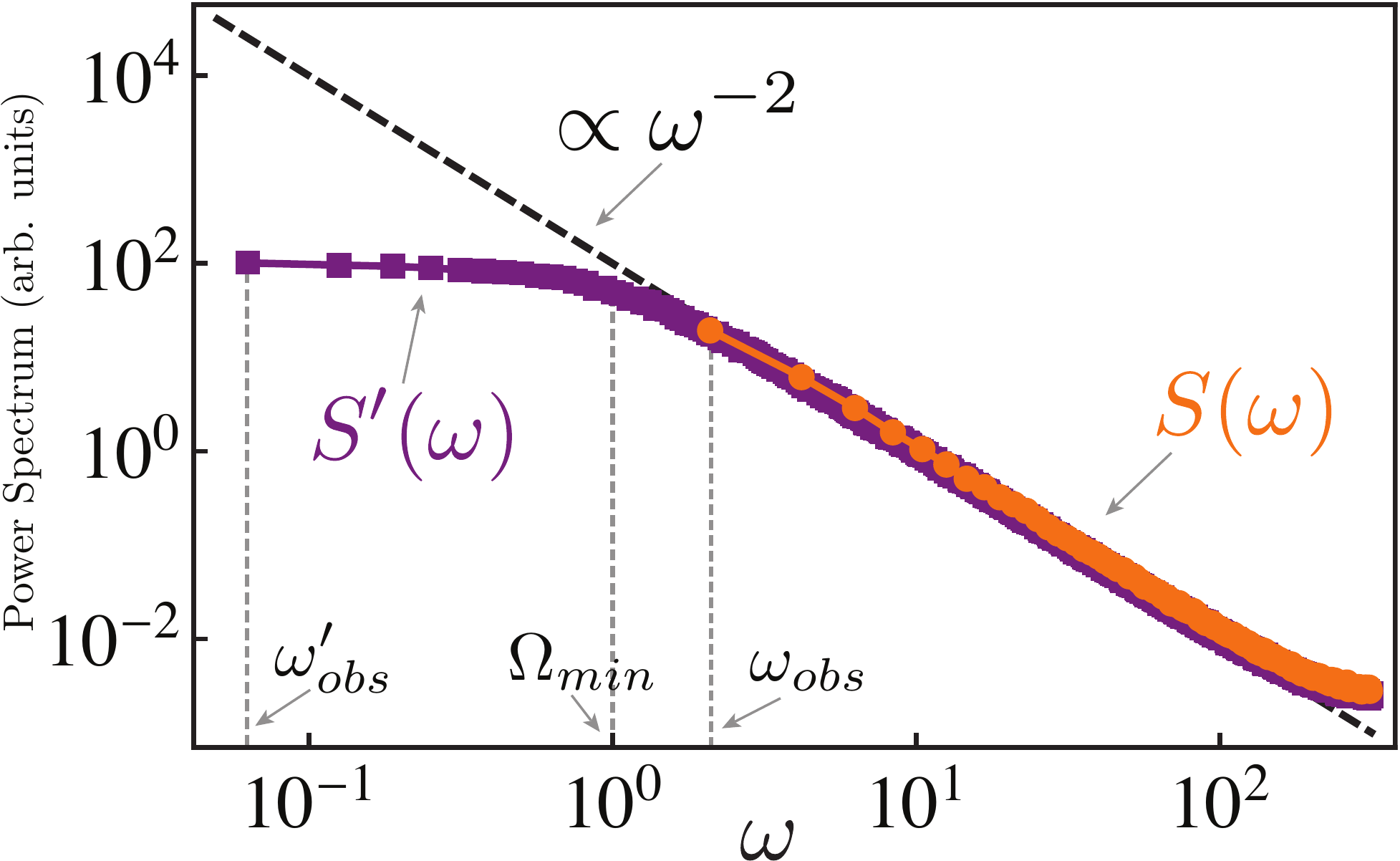}
\caption{(color online). Power spectral density crossover. Log-log plots of the power spectral density for the velocity of the archetypal one-dimensional Brownian motion simulated for two different (dimensionless) time intervals $T_{obs}$ and $T'_{obs}$, with $T_{obs} < T'_{obs}$. Simulations details are in the text. Orange dots: simulation run for $T_{obs}$ behaves as $1/\omega^2$ at low frequencies until $\omega_{obs}=2\pi/T_{obs}$. Purple squares: simulation run for a longer time $T'_{obs}$ behaves as $1/\omega^2$ down to $\approx\Omega_{min}$ and then changes to a different behavior until $\omega'_{obs}=2\pi/T'_{obs}$.}
\label{fig:Crossover}
\end{figure}

This, however, leads us to the opposite observation limit: $\Omega_{min}>\omega_{obs}$. Now $p(\Omega)$ can be resolved in $[0, \Omega_{min}]$ and it vanishes in the vicinity of the origin, as it should since there are no processes that have infinite relaxation times. Our main result \eqref{eq:1/f} then no longer holds and $S(\omega)$ has a functional form \emph{different} from $1/f$ at low frequencies. Nevertheless, $1/f$ noise may still be found in a band of larger frequencies and the power spectral density would exhibit a \emph{crossover} from $1/f$ to non-$1/f$---whose exact functional form would depend on $\bar{p}(\Omega)$---as frequency is swept toward the origin.

This is illustrated in Fig. \ref{fig:Crossover} for the archetypal Brownian motion in one dimension. A scaled and parameter-free version of the corresponding Langevin equation was simulated, namely $ \frac{d}{d\tau}\nu(\tau) + \nu(\tau) = \xi(\tau)$, with $\nu$ and $\tau$ being the dimensionless velocity and time, respectively. The dimensionless noise $\xi$ satisfies $\langle\xi(\tau)\rangle = 0$ and $\langle\xi(\tau)\xi(\tau')\rangle = \delta(\tau-\tau')$. In this scaled model, the minimum---and only---relaxation frequency is simply $\Omega_{min}=1$. For all the simulated realizations a time step of $0.01$ was used. Two sets of simulations were run. One set was simulated for $T_{obs} = 3$, the other for $T'_{obs} = 100$. The power spectral density of each set was computed from averaging over $1000$ realizations.

Since our derivation is quite general, the previous discussion can then be rephrased as a specific prediction. In any stochastic process, a $1/f$ power spectral density will be obtained down to the lowest measurable frequency $\omega_{obs}$ whenever $\omega_{obs}>\Omega_{min}$. Since the minimum relaxation frequency of a system is generally fixed, by increasing the observation time the reversed inequality, $\omega_{obs}<\Omega_{min}$, can be attained. As frequency is decreased to the new $\omega_{obs}$ the power spectral density now exhibits a crossover from $1/f$ to a different functional form. Indeed, as already mentioned a handful of experiments may have already shown such crossover \cite{ClarkeHsiang:1976,Verveen:1965,Verveen:1968,Gurp:1968,Clem:1981,Peng:1995,Anton:2012}, which we here
claim must be \emph{a universal feature of any stochastic process}.

%%%%%%%%%%%%%%%%%%%%%%%%%%%%%%%%%%%%%%%%
%%%%%%%%%%%%%%%%%%%%%%%%%%%%%%%%%%%%%%%%
\emph{Conclusions}.---
By realizing that initial conditions, hence non-stationarity, and the presence of relaxation processes are key, unavoidable features of any real system, we have presented a general derivation of $1/f$ noise that takes them into account. The ubiquitousness of these features allows us to apply our derivation to a broad class of systems and, at the same time, justify the ubiquitousness of $1/f$ noise itself.

Most importantly, we make a prediction related to the fact that the measured physical variables of most of the systems have a large number of degrees of freedom they can interact with, and hence are likely to have quite small relaxation frequencies. These would then usually require a very long observation time to be resolved. That is why we think many experiments do not report power spectral densities with a crossover from $1/f$ to non-$1/f$ behavior. However, such a crossover should \emph{always} be present, were such frequencies experimentally reachable, as we have numerically shown for the 1-D Brownian particle. Other ideal systems to look for such crossover are those with a very broad spectrum of relaxation times, such as glassy systems. An important aspect that emerges from our work is that, by taking advantage of this crossover in the power spectrum, the smallest relaxation frequency of the system can be experimentally measured. We thus hope our work will motivate experiments in this direction to clarify this very important and fundamental noise feature.

%%%%%%%%%%%%%%%%%%%%%%%%%%%%%%%%%%%%%%%%
%%%%%%%%%%%%%%%%%%%%%%%%%%%%%%%%%%%%%%%%

\begin{acknowledgments}
We thank Guy Cohen and Sebastiano Peotta for a critical reading of the manuscript. S.D. acknowledges partial support from the International Fulbright Science and Technology Award. M.D. acknowledges support from the Center for Magnetic Recording Research at UCSD.
\end{acknowledgments}

%%%%%%%%%%%%%%%%%%%%%%%%%%%%%%%%%%%%%%%%
%%%%%%%%%%%%%%%%%%%%%%%%%%%%%%%%%%%%%%%%

\bibliography{FlickerNoiseCrossover2}

\end{document}